\documentclass[letterpaper]{JHEP3}
\usepackage{array}

\usepackage{nicefrac}
\usepackage{amsmath}

\usepackage{amssymb,amsfonts,mathrsfs}
\usepackage{mathpazo}

\usepackage{graphicx}
\usepackage{epsfig}

\usepackage{amsfonts}

\renewcommand{\Im}{\mbox{Im}}

\newcommand{\Tr}{\mbox{Tr}}

\newcommand{\suup}{1}

\def\s{\sigma}

\def\t{\tau}
\def\a{\alpha}

\def\h{\eta}

\def\half{{\frac12}}

\def\IC{\relax\hbox{$\inbar\kern-.3em{\rm C}$}}

\def\IC{{\bf C}}

\def\bea{\begin{eqnarray}}
\def\eea{\end{eqnarray}}
\def\be{\begin{equation}}
\def\ee{\end{equation}}
\def\ba{\begin{align}}
\def\ea{\end{align}}
\def\bse{\begin{subequations}}
\def\ese{\end{subequations}}
\def\1F1{{}_1\!F_1}
\def\2F0{{}_2\!F_0}

\def\a{\alpha}

\def\h3{$\textrm{H}_3^+$}

\def\IC{{\mathbb C}}

\def\Tr{{\rm Tr}}

\def\lbldef#1#2{\expandafter\gdef\csname #1\endcsname {#2}}

\def\href#1#2{#2}

\newcommand{\beq}{\begin{equation}}
\newcommand{\eeq}{\end{equation}}
\newcommand{\ber}{\begin{eqnarray}}
\newcommand{\eer}{\end{eqnarray}}

\def\be{\begin{eqnarray}}
\def\ee{\end{eqnarray}}

\providecommand{\tabularnewline}{\\}

\def\({\left(}
\def\){\right)}
\def\[{\left[}
\def\]{\right]}
\def\<{\langle}
\def\>{\rangle}

\def\t{\tau}

\def\sixth{\frac16}

\title{On a modular property of ${\cal N}=2$ superconformal theories in four dimensions}
\author{
 Shlomo S. Razamat\\
\it Institute for Advanced Study, Princeton, NJ 08540, USA\\
}

\bigskip

\abstract{ 
In this note we discuss several properties of the Schur index of ${\cal N}=2$ superconformal theories
in four dimensions. In particular, we study modular properties of this index under
$SL(2,\, {\mathbb Z})$ transformations of its parameters. 

}

\begin{document}

\section{Introduction}

In two-dimensional conformal physics 
modular invariant partition functions have  
useful physical applications: for instance they encode the central charge of the theory and relate 
quantities computed in different regimes.
It is thus an  interesting question to ask whether there are partition functions of
higher-dimensional theories enjoying similar mathematical and physical properties. 
In this brief note we will discuss an example of such a partition function for four dimensional
${\cal N}=2$ superconformal theories.

A relatively simple quantity one can compute for conformal theories in four dimensions is
the superconformal index~\cite{Kinney:2005ej,Romelsberger:2005eg}, {\it i.e.} a partition function
on $S^3\times S^1$ with suitable boundary conditions for the fields. The superconformal index
of free theories, free chiral fields and free vector fields, can be expressed naturally in terms
of {\it elliptic Gamma functions}~\cite{Dolan:2008qi}. These special functions have very interesting and non trivial modular properties~\cite{SL3Za} which can be associated to an action of 
$SL(3,\,{\mathbb Z})\ltimes {\mathbb Z}^3$ on their parameters. 
Recently, in a very interesting paper~\cite{Spiridonov:2012ww} the authors studied such $SL(3,\,{\mathbb Z})\ltimes {\mathbb Z}^3$ transformations of indices of a family of
${\cal N}=1$  theories. It was observed in~\cite{Spiridonov:2012ww} that the modularly transformed
expressions for the index of this family of theories encode their  anomaly polynomials.\footnote{
See~\cite{Spiridonov:2009za,Total1,Total2} for previous work leading to this observation.} 
The expressions for the index, even for free fields,
 in terms of modularly transformed parameters 
have a rather different functional form from the index one starts with.
In particular, they do not have a direct interpretation as a usual superconformal index, {\it i.e.} a partition function on $S^3\times S^1$. A plausible speculation is that the transformed expression can be interpreted as a partition function of the theory on a different manifold and that the modular transformations thus relate partition functions on two different geometries:
 the precise details of such a claim, if true, are yet to be worked out.

In this note we will study  ${\cal N}=2$ superconformal theories and a special version of the 
${\cal N}=2$ superconformal index called the Schur index~\cite{Gadde:2011uv,Gadde:2011ik}.
 In this setup the non-trivial modular properties of the index tremendously simplify. In particular,
the modularly transformed expression of the index of a free hypermultiplet has functionally {\it the same}
form as the expression for the index one starts with. This means that even after performing the modular transformation the expressions might have a meaning as partition functions on $S^3\times S^1$.
The group associated to the modular properties of the Schur index is the simpler $SL(2, {\mathbb Z})$: {\it i.e.} the $SL(3, {\mathbb Z})$ transformations of the elliptic Gamma
functions reduce here to $SL(2, {\mathbb Z})$. More concretely, the Schur index is a function of 
at least one fugacity $q$ which can be conveniently parametrized as
\be
q=e^{2\pi i\,\tau}\,.
\ee
We will argue that the Schur index of a generic ${\cal N}=2$ superconformal theory
enjoys simple properties under the modular transformation $\tau\to -\frac1\tau$.
Moreover, we will show that the conformal central charge $c$ appears in a natural way
as an ingredient of the modular transformations of the Schur index.\footnote{The appearance 
of central charge $c$ here is the Schur index avatar of the anomaly polynomials discussed in~\cite{Spiridonov:2012ww}.} 
The parameter $\tau$ can be associated to the ratio of the radii of $S^1$ and $S^3$.
Thus, the modular properties we will discuss relate  $S^3\times S^1$
partition functions  with small radius of $S^1$ to the large radius limit. 

Before switching to the main part of the note let us mention several interesting contexts in which Schur
index makes an appearance:
{\it e.g.} it can be related to zero area~\cite{Gadde:2011ik} (and finite area~\cite{Tachikawa:2012wi})
 qYM in two dimensions as defined in~\cite{Aganagic:2004js};\footnote{
 Some of the modular properties of qYM are discussed in~\cite{Aganagic:2004js}.
} it preserves the same symmetries as 
a class of line operators~\cite{Gang:2012yr}; related to the latter it is sensible to define \textit{half} Schur index~\cite{Dimofte:2011py} by considering cutting the $S^3$ into two halves.  

\

This note is organized as follows. In section~\ref{lagsec} we introduce the Schur index 
and discuss its basic properties. Then we compute the Schur index for Lagrangian theories and explicitly
show the emergence of its elliptic and modular properties.
This discussion leads us to a generic formula for modular transformations 
of the Schur index~\eqref{boxed}. Next, in section~\ref{nonlagsec} 
we argue that the results obtained for Lagrangian theories imply simple modular properties  for, at least a class of, 
theories lacking a Lagrangian description.
We end with a brief discussion of our results in section~\ref{discsec}.
In appendix~\ref{sl3app} we discuss the relation of $SL(2,\,{\mathbb Z})$ transformations
of the Schur index to the $SL(3,\,{\mathbb Z})$ transformations of elliptic Gamma functions.
In appendix~\ref{appInt} we present several explicit expressions for Schur indices of Lagrangian
theories and show their relation to q-digamma function.
An additional appendix contains a technical result not essential to the main part of the note.

\

\

\section{Schur index and elliptic properties: Lagrangian theories}\label{lagsec}

The most general superconformal index one can write for
any ${\cal N}=2$ superconformal field theory in four dimensions depends on three 
fugacities coupled  to three different combinations of charges in the supeconformal
 algebra~\cite{Kinney:2005ej}. States contributing to such an index are annihilated by one of the eight 
supercharges (and its superconformal counterpart).  However, choosing to switch on less
fugacities one can count only states preserving more supersymmetry~\cite{Gadde:2011uv}.
In this note we consider the Schur index~\cite{Gadde:2011ik,Gadde:2011uv}  depending only on
one supeconformal fugacity $q$,
\be\label{traceI}
{\cal I}=\Tr(-1)^F\, q^{E-R}\,e^{-\beta_1(E+2j_1-2R-r)}\,e^{-\beta_2(E-2j_2-2R+r)}\, \prod_{i} { a}_i^{f_i}\,.
\ee
Throughout the paper we follow the notations of~\cite{Gadde:2011uv} and in particular $E$ is the conformal dimension, $j_{1,2}$
are the Cartans of $SU(2)_1\times SU(2)_2$ Lorentz isometry of $S^3$, $R$ is the Cartan of $SU(2)_R$ R-symmetry, and $r$ is the $U(1)_r$ R-charge.
The chemical potentials $\beta_{1,2}$ couple 
to the following combinations of bosonic charges
\be\label{susyC}
&&2\left\{{\cal Q}_{1+},\,{{\cal Q}_{1+}}^\dagger\right\}=E+2j_1-2R-r\equiv \delta_1\,,\\
&&2\left\{\widetilde{\cal Q}_{1\dot{-}},\,{\widetilde{\cal Q}}_{1\dot{-}}{ }^\dagger \right\}=E-2j_2-2R+r\equiv \delta_2\,,\nonumber
\ee where ${\cal Q}_{1+}$ and $\widetilde{\cal Q}_{1\dot{-}}$ have  charges
$$(E,j_1,j_2,R,r)=(\half,\half,0,\half,\half),\qquad (E,j_1,j_2,R,r)=(\half,0,-\half,\half,-\half)$$
respectively. 
Since both ${\cal Q}_{1+}$ and  ${\widetilde{\cal Q}}_{1\dot{-}}$ commute with $\delta_{1,2}$ and with
$E-R$ following the usual Witten index logic the 
 states contributing to the Schur index are annihilated by supercharges ${\cal Q}_{1+}$, ${\widetilde{\cal Q}}_{1\dot{-}}$ 
and their superconformal counterparts implying that the Schur index gets contributions only from states with $\delta_{1,2}=0$.
The Schur index is thus independent of $\beta_1$ and $\beta_2$ and we will omit these parameters from all the following expressions. 
Fugacities, 
\be{a}_\ell\equiv e^{2\pi i\,\a_\ell},\ee couple to global, flavor, charges $f_\ell$.
We will also define a modular parameter $\tau$ through,
\be
q\equiv e^{2\pi i\, \tau}\,.
\ee
An important assumption one makes when writing~\eqref{traceI} is that
\be
\left|q\right|<1\,,\qquad \Im\{\tau\}> 0\,,
\ee and in particular the index~\eqref{traceI} is to be thought of as an expansion in $q$ around $q=0$.
The Schur index is guaranteed to have a well defined $q$-expansion 
for any ${\cal N}=2$ superconformal-theory\footnote{Assuming that there is a finite number of protected states with given $E-R$ quantum numbers.}~\cite{Gadde:2011uv}
following from the fact that
\be\label{ER}
E-R=\frac{1}{4}\,\left(\delta_1+\delta_2+\delta_3+\delta_4\right)\,,
\ee with
\be\label{susyC2}
&&2\left\{{\cal Q}_{1-},\,{{\cal Q}_{1-}}^\dagger\right\}=E-2j_1-2R-r\equiv\delta_3\,,\\
&&2\left\{\widetilde{\cal Q}_{2\dot{+}},\,{\widetilde{\cal Q}}_{2\dot{+}}{ }^\dagger \right\}=E+2j_2+2R+r\equiv \delta_4\,.\nonumber
\ee Here  ${\cal Q}_{1-}$ and $\widetilde{\cal Q}_{2\dot{+}}$ have the charges
$$(E,j_1,j_2,R,r)=(\half,-\half,0,\half,\half),\qquad (E,j_1,j_2,R,r)=(\half,0,\half,-\half,-\half)$$
respectively. The facts~\eqref{ER} and~\eqref{susyC2} imply that all states contributing to the Schur index have non-negative $E-R$ charge. 

Another assumption one usually makes is that
\be
|a_\ell|=1\,,\qquad \Im\{\a_\ell\}=0\,.
\ee 
For instance the representation of the index as a sum over orthogonal functions in~\cite{Gadde:2011ik,Gadde:2011uv,GRR} is strictly valid only with such an assumption.
However, it is often useful to {\it analytically continue} the expressions for the index to complex values of $\a_\ell$. It is important to remember that the analytically continued expressions do not a priori
have a physical interpretation as a trace over the Hilbert space of the form~\eqref{traceI}.
However,  in certain cases such an interpretation might be available.
For example, in~\cite{GRR} some analytical properties of indices in flavor fugacities were related to indices of IR theories 
in presence of vacuum expectation values for gauge invariant operators. 
In this paper we will also complexify $\a_\ell$ by analtyic continuation
of the expressions for the index.

Let us discuss the basic building blocks of the Lagrangian ${\cal N}=2$ theories: the free hypermultiplet and the free ${\cal N}=2$ vector multiplet.
The Schur index of a free hypermultiplet is given by~\cite{Gadde:2011uv,Gadde:2011ik}
\be\label{freeH}
{\cal I}_H(a;\,q)=\prod_{\ell=0}^\infty \frac{1}{1-q^{\half+\ell} \,{a}^{} }\,\frac{1}{1-q^{\half+\ell} \,{a}^{-1} }\equiv
\prod_{\ell=0}^\infty \frac{1}{1-q^{\half+\ell} \,{a}^{\pm1}}
=\frac{1}{\theta(q^\half { a}^{};q)} \,.
\ee
The only fields of the hypermultiplet contributing to the index are two scalars, $Q$ and $\widetilde Q$,
together with a certain derivatives of those~\cite{Gadde:2011uv,Gadde:2011ik}, which we will denote by $\partial$ (see table~\ref{letters} for a list of ``letters'' contributing to the Schur index). The weights of $Q$ and $\widetilde Q$ are 
$q^\half\,a$ and $q^\half\,a^{-1}$ respectively; the derivative $\partial$ contributes a factor of $q$.
 The fugacity ${a}$ couples to $U(1)_f$  flavor symmetry giving the two half-hypermultiplets opposite charges.
 We use the usual shorthand notation where the ambiguous signs in arguments of functions mean that we have a product 
over the choices of the signs. The theta function and the Pochhammer symbol are defined as 
\be
&&\theta(z;\,q)=\prod_{i=0}^\infty (1-z\,q^i)\,(1-z^{-1}\,q^{i+1})\,,\\
&&(a;\,q)=\prod_{i=0}^\infty (1-a\,q^i)\,.\nonumber
\ee
 Note that the Schur index of the free hypermultiplet is trivially invariant 
under the following transformation of the chemical potential $\a$ 
\be
\a\to \a+1\,.
\ee Moreover it has simple properties under
\be\label{elltr}
\a \to \a+\tau\,.
\ee 
Making such a transformation of the flavor chemical potential we are implicitly complexifying it.
We can use the following property of the theta function
\be\label{tr}
\theta(q^rz;q)=(-1)^r\,z^{-r}\,q^{-\half r(r-1)}\,\theta(z;q)\,,\qquad r\in {\mathbb Z}\,,
\ee to write 
\be\label{free2}
{\cal I}_H(a;\,q)\quad \underset{{  a}\to q\,{  a}}{\longrightarrow}\quad 
-q^{\half}\,{  a}\, \frac{1}{\theta(q^\half {  a};\,q)}
=-q^{\half}\,{  a}\,{\cal I}_H(a;\,q)\,.
\ee  Here a subtlety with a physical interpretation of the analytically continued expression arises.
Note that after the transformation~\eqref{elltr} the expression for the index~\eqref{free2} does not have clear cut physical meaning as a trace over the states: {\it e.g.} the expansion of the index in powers of $q$ does not contain $q^0=1$ term which would correspond to the neutral vacuum of the theory. The transformation~\eqref{elltr}
is equivalent to $a\to q\,a$, which might have  been interpreted    as weighing the states with 
$q^{E-R+f}$ and not with $q^{E-R}$ in~\eqref{traceI}. However a trace over the Hilbert space
with weight $q^{E-R+f}$ diverges for the free hypermultiplet: there are infinite number of states 
with the same $E-R+f$ quantum numbers: {\it e.g.} the  operators $\left(\widetilde Q\; \partial \widetilde Q\right)^\ell$ have $E-R+f=0$ for any $\ell$. Mathematically the discrepancy between 
the finite expression~\eqref{free2} and the infinite trace interpretation can be stated as the fact that
the expansion
\be
\frac{1}{1-q^{\half}{  a}^{-1}}=\sum_{\ell=0}^\infty q^{\half \ell}\,{  a}^{-\ell}\,,
\ee and the transformation~\eqref{elltr} do not commute. The  divergent 
trace interpretation is obtained by first doing the expansion and then transforming, and
the mathematically finite  expression~\eqref{free2} is obtained by reversing the order of the two operations. Thus, to summarize, the transformation~\eqref{elltr} is a property of the 
analytically continued expression for the index, which  does not have a physical interpretation
as a sum over the states but only as an analytic continuation of the partition function on $S^3\times S^1$. 

\

The index of the free hypermultiplet has another very interesting property: it transforms naturally
under a modular transformation,
\be
q\to q'\equiv e^{-\frac{2\pi i}\tau}\,,\qquad a\to a'\equiv e^{\frac{2\pi i \a}{\tau}} \,.
\ee Under this modular transformation the theta functions 
and the $\eta$-function transform
 in the standard way
\be\label{modP0}
&&-i\,e^{\pi i(\frac{\a^2}{\tau}+\frac16(\tau+\frac1\tau)+\a(\frac1\tau-1))}\,\theta(a;q)=\theta(a';q')\,,\\
&&(-i\,\tau)^{\half}\,q^{\frac{1}{24}}(q;q)=q^{'\frac{1}{24}}(q';q')\,.\nonumber
\ee
Parametrizing the flavor fugacity as $a\equiv -b\equiv -e^{2\pi i \eta}$ this implies that
\be\label{freemod}
q^{\frac{1}{24}}\,{\cal I}_H(-b;\,q)=e^{\frac{\pi i}{\tau}\eta^2}\;{q'}^{\frac{1}{24}}\,{\cal I}_H(-b';\,q')\,.
\ee Note that under~\eqref{modP0} $-1$ in the argument of the theta function transforms to $q^\half$
and $q^\half$ transforms into $-1$: that is the reason to choosing $a\equiv-b$ for this example.
Thus, keeping in mind the modular transformations it will be more natural to define a {\it modified} Schur index for Lagrangian theories,
\be\label{twisted}
{\cal I}^T=\Tr(-1)^{F+B}\,q^{E-R}\,\prod_{i}a_i^{f_i}\,.
\ee Here $B$ is the baryon number giving charge $\pm1$ to the two half-hypermultiplets $Q$
and $\widetilde Q$.
The charge $B$ is part of the flavor symmetry and $(-1)^B$ 
can be absorbed in flavor fugacities $a_i^{f_i}$ and thus the indices~\eqref{twisted} and~\eqref{traceI}
are equivalent under redefinition of flavor fugacities.
 We note in passing that 
\be
\frac{1}{24}=\frac{c_H}{2}\,,
\ee where $c_H=\frac{1}{12}$ is the $c$ conformal central charge of a free hypermultiplet.

\

Let us now remind the reader how  gauging of a symmetry affects the index.
The Schur index of the vector multiplet is given by
\be
{\cal I}_V=\frac{(q;q)^{2N-2}}{N!\Delta({\mathbf z})}\prod_{i\neq j}\theta(z_i/z_j;q)\,,
\ee where $\Delta(z)$ is the $SU(N)$ Haar measure,\footnote{
For sake of simplicity we restrict ourselves in this note to $SU(N)$ gauge theories.}
\be
\Delta({\mathbf z})=\frac{1}{N!}\prod_{i\neq j}(1-z_i/z_j)\,.
\ee By the usual rules of computing the index when an $SU(N)$  symmetry is gauged one computes the index by adding a vector multiplet and projecting on gauge invariant states.
Thus, given the index ${\cal I}(\mathbf z)$ of some  theory with $SU(N)_{\mathbf z}$ flavor symmetry the index of the theory with this symmetry gauged is given by
\be
\frac{(q;q)^{2N-2}}{N!}\oint \prod_{i=1}^{N-1}\frac{dz_i}{2\pi i\,z_i}\,\prod_{i\neq j}\theta(z_i/z_j;q)\,{\cal I}(\mathbf z)\,.
\ee
Let us now take a general conformal theory which contains $2N$ fundamental hypermultiplets of gauged $SU(N)$ symmetry which might be coupled to other matter (neutral under this gauge symmetry) through gauge interactions. The modified index
of such a theory is given by\footnote{The integrals involved in the 
computations of indices of Lagrangian theories can in general  be explicitly performed 
and the result can be formulated in terms of special functions. In appendix~\ref{appInt} 
we give several examples.
} 
\be
&&{\cal I}_A^T=\\
&&\qquad \oint\cdots \left[
\frac{(q;q)^{2N-2}}{N!}\oint \prod_{i=1}^{N-1}\frac{dz_i}{2\pi i\,z_i}\,\prod_{i\neq j}\theta(z_i/z_j;q)\,
\prod_{i=1}^N\prod_{j=1}^{2N}\frac{1}{\theta(-q^\half\, z_i\,x_j;\,q)}
\right]\equiv\oint\cdots {\cal I}^T_{B}\,.\nonumber
\ee Here ellipses represent additional matter and the first contour integral represents schematically 
any additional gaugings one might have. 
The quantity,
 \be \prod_{i=1}^N\prod_{j=1}^{2N}\frac{1}{\theta(-q^\half\, z_i\,x_j;\,q)}, \ee is the
modified index of the $2N$  hypermultiplets with $x_i$ labelling a $U(2N)$ symmetry rotating them.\footnote{This symmetry might be broken to its smaller sub-group but this is not essential for our argument.}
First we note that the integrand of $\oint \prod_{i=1}^{N-1}\frac{dz_i}{2\pi i\,z_i}$ in ${\cal I}_B^T$ is a doubly periodic function of $\zeta_i$,
\be
\zeta_i\sim\zeta_i+1\,, \qquad \zeta_i\sim\zeta_i+\tau\,,
\ee where $z_i=e^{2\pi i \zeta_i}$. This means that the chemical potentials $\zeta_i$ naturally live on a torus with modular 
parameter $\tau$.
\begin{figure}[htbp]
\begin{center}
\begin{tabular}{c}
\includegraphics[scale=0.4]{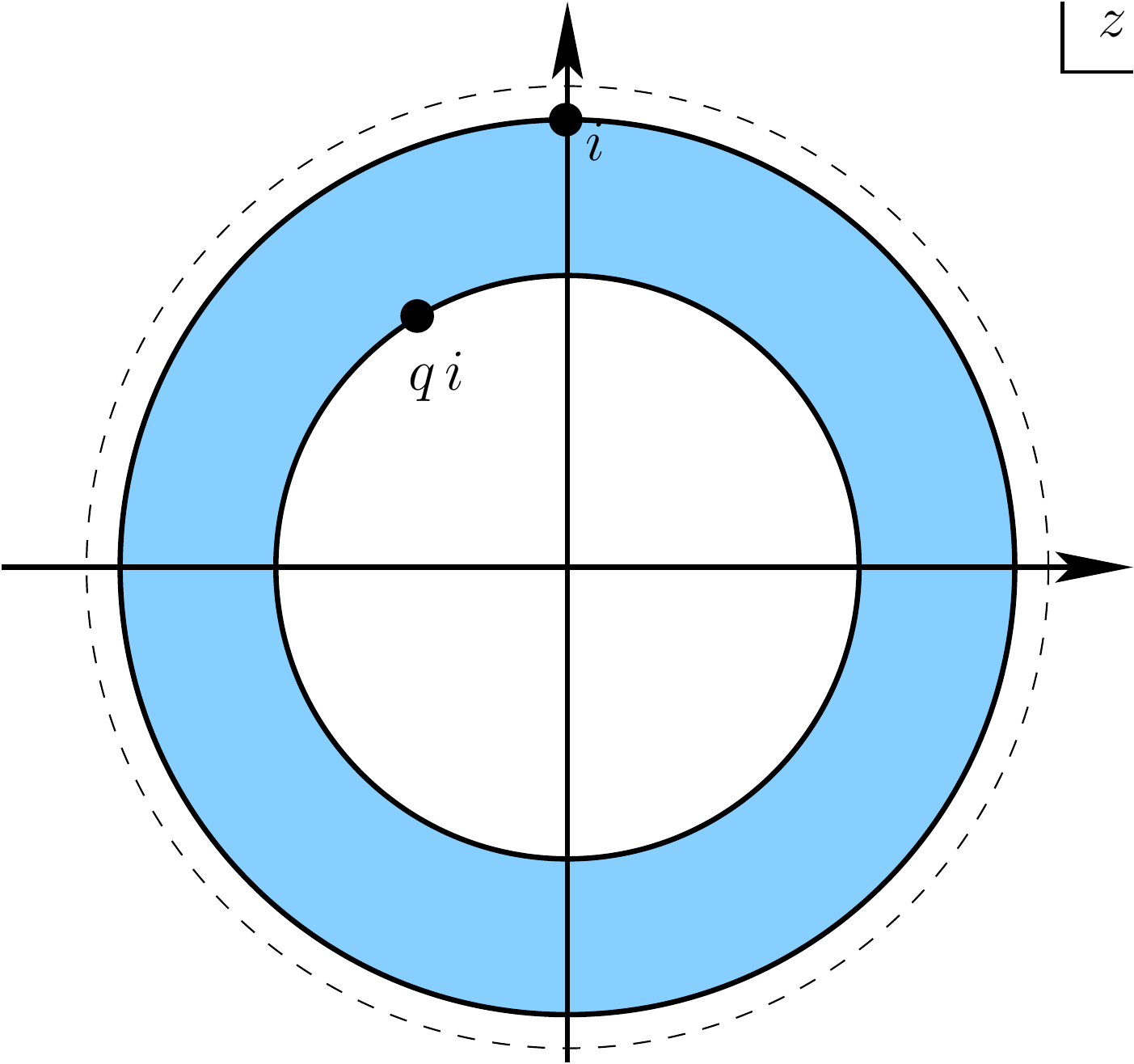}
\end{tabular}
\end{center}
\caption{The torus on the (complexified) flavor fugacity plane $z=e^{2\pi i \, \zeta}$. The two sides of the annulus are to be identified, $z\sim q\,z$.
The dashed line is the unit circle: this is the ${\cal A}$-cycle of the torus around which the flavor symmetry is integrated over when it is being gauged.\label{tor}}
\end{figure}
This torus is an annulus in fugacity space depicted in figure~\ref{tor}. A nice consistency check 
of this picture is that the sum over residues of the integral ${\cal I}^T_B$ in each annulus
($k$th annulus is bounded  by $z\,q^k$ and $z\,q^{k+1}$ with $z$ on the unit circle) vanishes
making the integration contour rigidly contractable to any vicinity of the origin (which is an accumulation point of poles).\footnote{For the integrand to be
 an elliptic function it is crucial that the number of hypermultiplets is twice the number of colors:
{\it i.e.} the elliptic properties of the Schur index here are a signature of conformal invariance of the theory.} 

\

Let us now discuss how ${\cal I}_B$ transforms under 
the modular transformation~\eqref{modP0},
\be
{\cal I}^T_B({\bf x};q)&=&
\frac{(q;q)^{2N-2}}{N!}\oint \prod_{i=1}^{N-1}\frac{dz_i}{2\pi i\,z_i}\,\prod_{i\neq j}\theta(z_i/z_j;q)\,
\prod_{i=1}^N\prod_{j=1}^{2N}\frac{1}{\theta(-q^\half\, z_i\,x_j;\,q)}\\
&=&
\frac{1}{N!}\left(\frac{q'}{q}\right)^{\frac{2N^2-N}{12}}\,
\left[\frac{i(q';q')^2}{\tau}\right]^{N-1}\left(\frac{q'}{q}\right)^{\frac{N-1}{12}}\,
e^{\frac{\pi i\,N}{\tau}\,\sum_{i=1}^{2N}\,\eta_i^2}\,i^{N(N-1)}\,
\times\nonumber\\
&&\qquad\qquad\int_{1}^{{q'}^{}}\prod_{i=1}^{N-1}\frac{\tau dz'_i}{2\pi i z'_i}\,\prod_{i\neq j}\theta(z'_i/z'_j;q')\,
\prod_{i=1}^N\prod_{j=1}^{2N}\frac{1}{\theta(-{q'}^\half\, z'_i\,x'_j;\,q')}\nonumber\\
&\equiv&i^{N^2-1}\,e^{\frac{\pi i\,N}{\tau}\,\sum_{i=1}^{2N}\,\eta_i^2}\,\left(\frac{q'}{q}\right)^{\frac{2N^2-1}{12}}\,
{\cal I'}^T_B({\bf x}';q')\nonumber\\
&=&i^{N^2-1}\,
e^{\frac{\pi i\,N}{\tau}\,\sum_{i=1}^{2N}\,\eta_i^2}\,\left(\frac{q'}{q}\right)^{\frac{2N^2\,c_H+(N^2-1)\,c_V}{2}}\,
{\cal I'}^T_B({\bf x}';q')\,.\nonumber
\ee
\begin{figure}
\begin{center}
\begin{tabular}{c}
\includegraphics[scale=0.6]{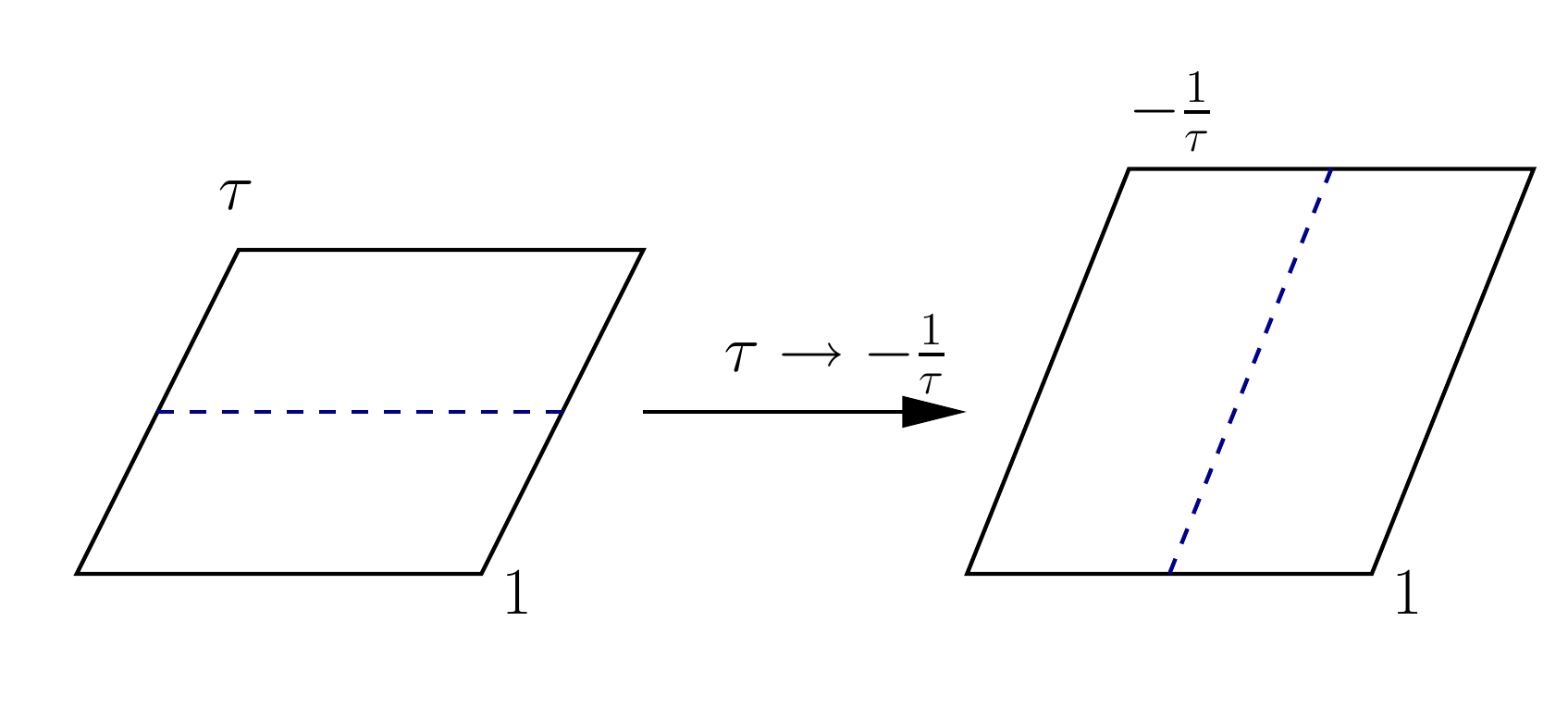}
\end{tabular}
\end{center}
\caption{The modular transformation interchanges the integrations over the two cycles: before 
performing modular transformation the integration is along the ${\cal A}$-cycle of the torus in the 
gauge chemical potential complex plane and after the transformation the integration is along the ${\cal B}$-cycle of the dual torus.
The dashed lines represent integration contours.
\label{mod}}
\end{figure}
Here,
\be
x_i\equiv e^{2\pi i\,\eta_i}\,,\qquad c_V=\frac16\,.
\ee 
After modular transformation the integrations are along an open contour running from ${q'}^k$  to ${q'}^{k+1}$
and parametrized by $e^{-\frac{2\pi i \,k\,\zeta}{\tau}}$ for $\zeta\in[0,1]$ for arbitrary integer $k$.
The $z_i$ integrations are along the ${\cal A}$-cycle of the torus: $z=e^{2\pi i \zeta_i}$ and then
$\oint\frac{dz_i}{2\pi i z_i}= \int_0^1d\zeta_i\equiv \int_{\cal A} d\zeta_i$.
 However, after the modular transformation the integration
is around the ${\cal B}$-cycle of the dual torus:
 $\int_{1}^{{q'}^{}} \frac{dz'_i}{2\pi i z'_i}$ is $\int_{\cal B} d\zeta'_i$ (as illustrated in figure~\ref{mod}).

Thus, we see that ${\cal I}^T_B$ has simple modular transformation.
Note that the ($\zeta_i$-dependent) integrand of  ${\cal I'}^T_B(\;;q')$ is exactly the same as for ${\cal I}^T_B(\;;q)$ 
upon changing the parameters to the modularly transformed ones. In particular,
all the $\zeta_i$-dependent phases, coming from the modular transformation of theta functions, cancel out: to achieve this the fact that the number
 of flavors is twice the number of colors is essential and thus this property is a consequence of conformal invariance.\footnote{In the context of~\cite{Spiridonov:2012ww}
 the absence of such phases was interpreted as an implication
of the absence of gauge anomalies.}
Moreover, the phase factor depending on flavor chemical potentials is the same as for $N$ fundamental
hypermultiplets of $U(2N)$. This implies that the procedure above can be re-iterated: modified
indices of any Lagrangian theory have simple modular properties.
Setting all flavor fugacities to one we can write for a generic Lagrangian theory,
\be\label{boxed}
\boxed{
q^{\half\,c}{\cal I}^T(1;\,q)=
i^{n_V}\,{q'}^{\half\,c}{\cal I'}^T(1;\,q')}\,,
\ee where $n_V$ is the number of vector multiplets.\footnote{
Note that $n_V$ is given by the following relation to the central charges $\frac14 n_V=2a-c$,
see {\it e.g.}~\cite{Shapere:2008zf}.
}

To summarize, it is natural to multiply the modified index with $q^{\half \, c}$: with this normalization the index has nice properties under modular transformations.
Taking $\tau\equiv i\beta \to 0$ ($q\to 1$) {\it naively } the modified index scales as\footnote{Note that in this limit the modified index vanishes. 
The modified index counts the difference between protected states with even/odd $B$ quantum numbers and this difference thus exponentially decreases for states 
with large charges.
} 
\be
{\cal I}^T\sim e^{-\frac{\pi \,c}{\beta}}\,.
\ee This result is true for free fields but it has to be corrected for the interacting theories\footnote{
Interacting theories here include also zero-coupling limit of gauge theories since the projection to gauge singlets changes the assymptotic behavior of the index.
} since the integrals along the ${\cal B}$-cycles involved in the definition
of the index have singular behavior in this limit: 
thus there is no robust statement about $c$ capturing a scaling of the index in the $q\to1$ limit.

\

\

\begin{table}
\begin{centering}
\begin{tabular}{|c|r|r|r|r|r|r|c|c|}
\hline
Letters & $  E$ & $j_1$ & $  j_2$ & $R$ & $r$ & $B$ &  $\mathcal{I}(q)$
&  $\mathcal{I}^T( q)$ \tabularnewline
  \hline
   \hline
$  \lambda_{\suup-}$ & $  \frac{3}{2}$ & $  -  \frac{1}{2}$ & $0$ & $  \frac{1}{2}$ & $-  \frac{1}{2}$ & $0$ &  $-q$ & $-q$ \tabularnewline
  \hline
$  \bar{\lambda}_{\suup\dot{+}}$  & $  \frac{3}{2}$ & $0$ & $  \frac{1}{2}$ & $  \frac{1}{2}$ & $  \frac{1}{2}$ & $0$&   $-q$ & $-q$ \tabularnewline
  \hline
\hline
$Q$ & $1$ & $0$ & $0$ & $  \frac{1}{2}$ & $0$  & $1$ &  $q^\half$ &  $-q^\half$\tabularnewline
  \hline
${\widetilde Q}$ & $1$ & $0$ & $0$ & $  \frac{1}{2}$ & $0$  & $-1$ &  $q^\half$ &  $-q^\half$\tabularnewline
  \hline
    \hline
$  \partial\; (\equiv\partial_{-\dot{+}})$ & $1$ & $  -  \frac{1}{2}$ & $  \frac{1}{2}$ & $0$ & $0$    & $0$ &  $q$ & $q$ \tabularnewline
\hline
\end{tabular}
\par  \end{centering}
  \caption{Contributions to the index from  ``single letters'' of free fields.
  We denote by $(\phi, \bar \phi,  \lambda_{I,\alpha}, \bar\lambda_{I\,\dot \alpha},  F_{\alpha \beta}, \bar F_{\dot \alpha \dot \beta})$
 the components of the adjoint ${\cal N} = 2$ vector multiplet,  by $(Q, \bar Q, \psi_\alpha, \bar \psi_{\dot \alpha})$ the
 components  of the  ${\cal N} = 1$
chiral multiplet (two chiral multiplets $Q$ and $\widetilde Q$ form a hypermultiplet),  and by $\partial_{\alpha \dot \alpha}$ the spacetime derivatives.
}
\label{letters}
\end{table}
Let us now give a huristic physical explanation of the modular property of the modified Schur
 index.\footnote{We thank C.~Beem, A.~Gadde, T.~Dimofte, and L.~Rastelli for important comments
very relevant for this discussion.}
Keeping in mind that the states contributing to the index have $\delta_{1,2}=0$ 
one can write the Schur index~\eqref{traceI} as
\be
{\cal I}=\Tr(-1)^F\, q^{\half\left(E+j_2-j_1\right)}\,.
\ee 
Moreover, note that for states contributing to Schur index\footnote{Note that 
 $(-1)^{F+B}=(-1)^{2R}$ holds without restricting to the states contributing to the Schur index.
We can then use the relation $E-j_2+j_1=2R$ holding for states contributing to Schur index to write~\eqref{FpB}.
 We thank T.~Dimofte for this observation.}
\be\label{FpB}
(-1)^{F+B}= (-1)^{E-j_2+j_1}=(-1)^{-E+j_2-j_1}\,.
\ee  
Thus, the {\it modified} Schur index can be written as
\be\label{indSymm}
{\cal I}^T=\Tr\,(-1)^{|E-(j_2-j_1)|}\, q^{\half\left(E+j_2-j_1\right)}\,.
\ee 
We see that in the computation of the modified Schur index there are two circles 
appearing on the same footing\footnote{Modulo the fact that 
the trace itself breaks the symmetry between $S^3$ and $S^1$.}:
 the temporal circle corresponding to  $E$ and the circle on $S^3$  
corresponding to  $j_2-j_1$. These two circles form the geometric torus responsible 
for the modular properties of the index. Taking $\tau$ to be $i\beta$, the  real parameter $\beta$ 
can be thought of as the ratio of the radii of the two circles.  
For index of free theories not refined with flavor fugacities 
the two circles can be swapped leading to its symmetry under modular transformations.
For interacting theories (or refining with flavors fugacities)
 the symmetry between the two circles is broken by the introduction
of holonomies for the gauge (flavor) fields around the temporal circle:
 exchanging the two circles by modular transformation we  have thus to introduce the holonomies around the spatial circle.

\

\

\

\section{Strongly-coupled theories: example of $E_6$ SCFT}\label{nonlagsec}

\

Let us try to extend the analysis of the previous section to ${\cal N}=2$ theories which
do not have a Lagrangian description.
 A large number of such theories belongs to the so called class ${\cal S}$~\cite{Gaiotto:2009we, Gaiotto:2009hg}.
Theories in class ${\cal S}$ are interconnected by webs of S-dualities. In particular some of the 
strongly interacting non-Lagrangian SCFTs can be incorporated into a bigger theory
by adding additional matter, and such a bigger theory in certain cases possesses dual Lagrangian 
description. These dualities can be used to deduce the index of the strongly-coupled theories, 
{\it e.g.} as was done in~\cite{Gadde:2010te} for the $E_6$ SCFT~\cite{Minahan:1996fg}. In this section we discuss the modular properties
of the Schur index of the $E_6$ SCFT. We expect that other strongly-coupled theories of class ${\cal S}$ 
will have similar properties.

\subsection*{Sphere with two maximal and one minimal punctures}

The theories of class ${\cal S}$ are associated to Riemann surfaces~\cite{Gaiotto:2009we} and we assume here familiarity with the jargon of this correspondence.
We start with the free building blocks of theories of class ${\cal S}$ 
corresponding to a sphere with two maximal and one minimal puncture: here we will
consider the generic $SU(N)$ case.
This building block is a free hypermultiplet in bi-fundamental representation of $SU(N)\times SU(N)$.
The modified Schur index is defined by
\be\label{suNind}
{\cal I}^T(a,{\bf x}, {\bf y};q)=\frac{1}
{\prod_{i,j=1}^N\theta(-q^\half x_iy_j a;q)}
\,.
\ee Here we take 
\be
\prod_{i=1}^Nx_i=\prod_{i=1}^Ny_i=1\,.
\ee 
We can glue two free bi-fundamental hypermultiplets together to form the $SU(N)$ $N_f=2N$ SCFT. 
The Riemann surface will be a sphere with two maximal and two minimal punctures. Following the 
results of the previous section we write
\be
{\cal I}^T(a,b,{\bf x},{\bf y};q)&=&\frac{(q;q)^{2(N-1)}}{N!}\oint\prod_{i=1}^{N-1}\frac{dz_i}{2\pi i z_i}\,
\prod_{i\neq j}\theta(z_i/z_j;q)
\,{\cal I}^T(a,{\bf x},{\bf z};q){\cal I}^T(b,{\bf y},{\bf z}^{-1};q)\nonumber\\
&\equiv&i^{N^2-1}\,\left(\frac{q'}{q}\right)^{\frac{2N^2-1}{12}}\,
e^{\frac{\pi i N}{\tau}(N(\a^2+\beta^2)+\sum_{i=1}^N(\eta_i^2+\gamma_i^{2}))}\,
{{\cal I}'}^T(a',b',{\bf x}',{\bf y}';q')\,.
\ee Here
\be
a=e^{2\pi i \a}\,,\qquad
b=e^{2\pi i \beta}\,,\qquad
x_i=e^{2\pi i \eta_i}\,,\qquad
y_i=e^{2\pi i \gamma_i}\,.
\ee

\subsection*{Sphere with three maximal punctures}
Here we specialize to the sphere with three maximal punctures of $SU(3)$: the $E_6$ SCFT~\cite{Minahan:1996fg}.
The index of this theory has a nice form as a sum over irreps of $SU(3)$~\cite{Gadde:2011uv,Gadde:2011ik}, which makes the connection to~\cite{Aganagic:2004js} clear.
However, an expression for the index which we find to be more convenient for checking its modular properties is 
the expresion in terms of a contour integral derived in~\cite{Gadde:2010te}.
Specializing to the Schur case, this integral expression for the index reduces to a finite sum over residues and is given in $SU(3)^2\times U(1)\times U(1)$
covariant form as,\footnote{
In the Schur limit the expression of~\cite{Gadde:2010te} tremendously simplifies:
The integral term in eq.~(3.19) in~\cite{Gadde:2010te} vanishes identically.  
The reason is that the contour integral of eq.~(3.18) in that paper is pinched 
by pairs of poles in the Schur limit and
at the same time there is a multiplicative vanishing contribution. Thus, 
in the Schur limit only the poles which pinch the contour of integration give finite contribution.
This relation between the index of $N_f=6$ $SU(3)$ SYM and the $E_6$ SCFT also implies an identity 
between Schur polynomials and theta functions which we present in appendix~\ref{schurapp}.
}
\be\label{e6dif}
&&{\cal I}_{E_6}({\bf z}=(wr,w^{-1}r,r^{-2}),{\bf x},{\bf y};\,q)=\\&&\qquad\qquad
\frac1{\theta(w^{-2};q)}\;\left.{\cal I}(\frac{w^{\frac{1}{3}}}{r},\frac{w^{-\frac{1}{3}}}{r};{\bf x},{\bf y})\right|_{w\to q^{-\half}\,w}+
\frac1{\theta(w^{2};q)}\;\left.{\cal I}(\frac{w^{\frac{1}{3}}}{r},\frac{w^{-\frac{1}{3}}}{r};{\bf x},{\bf y})\right|_{w\to q^{\half}w}\,.\nonumber
\ee
On the left-hand-side of~\eqref{e6dif} the $U(1)$ fugacities are combined into $SU(3)$
 ones.

Let us mention several facts about the index of the four-punctured sphere with two maximal and
two minimal punctures~\cite{Gadde:2010te}. First, because of S-duality we have,
\be\label{first}
{\cal I}(a,b,{\bf x},{\bf y})={\cal I}(b,a,{\bf x},{\bf y})\,.
\ee Second, 
\be\label{second}
{\cal I}(a,b,{\bf x},{\bf y})={\cal I}(e^{\frac{2\pi i}3\,\ell}a,^{\frac{-2\pi i}3\,\ell}b,{\bf x},{\bf y})\,,\qquad
\ell\in {\mathbb Z}\,.
\ee This can be also phrased as the fact that the index has an expansion in integer powers of fugacities 
 $w$ 
and $r$ appearing in~\eqref{e6dif} as a consequence of Argyres-Seiberg duality~\cite{argyres-2007-0712}.

We are now ready to tackle the modular transformation of the (modified) Schur index of the $E_6$
SCFT. A priori it is not clear what the analogue of the modified index is for the strongly coupled SCFT.\footnote{
A natural definition of the modified index would be through~\eqref{indSymm}.
}
However, since this theory has no natural $U(1)$ flavor symmetry to play the role of the baryonic symmetry we modify the index with, one can assume that the modified index of $E_6$ SCFT coincide with the Schur
index; and indeed we will see that this is a consistent assumption. 
We can write the following
\be
{\cal I}(\frac{w^{\frac{1}{3}}}{r},\frac{w^{-\frac{1}{3}}}{r};{\bf x},{\bf y})&=&
{\cal I}^T(\frac{-w^{\frac{1}{3}}}{r},\frac{-w^{-\frac{1}{3}}}{r};{\bf x},{\bf y})=
{\cal I}^T(\frac{e^{-\frac{\pi i }{3}}w^{\frac{1}{3}}}{r},\frac{e^{\frac{\pi i}{3}}w^{-\frac{1}{3}}}{r};{\bf x},{\bf y})=\\
&=&{\cal I}^T(\frac{e^{\frac{\pi i }{3}}w^{\frac{1}{3}}}{r},\frac{e^{-\frac{\pi i}{3}}w^{-\frac{1}{3}}}{r};{\bf x},{\bf y})\,.\nonumber
\ee
 The index of the $E_6$ SCFT is given thus by
\be
&&{\cal I}_{E_6}({\bf z},{\bf x},{\bf y};q)=
\frac{1}
{\theta(w^{-2};q)}\,
{\cal I}^T(e^{\frac{\pi i}{3}}q^{-\sixth}\frac{w^{\frac{1}{3}}}{r},e^{-\frac{\pi i}{3}}q^{\sixth}\frac1{w^{\frac{1}{3}}r},{\bf x},{\bf y})+\nonumber\\
&&\qquad \qquad\qquad\qquad\frac{1}
{\theta(w^{2};q)}\,\,
{\cal I}^T(e^{-\frac{\pi i}{3}}q^{\sixth}
\frac{w^{\frac{1}{3}}}{r},e^{\frac{\pi i}{3}}q^{-\sixth}
\frac1{w^{\frac{1}{3}}r},{\bf x},{\bf y})\,.\nonumber
\ee
Under modular transformation we get
\be
&&{\cal I}_{E_6}({\bf z},{\bf x},{\bf y};q)=
 i\,e^{\frac{3\pi i \sum_{i=1}^N(\eta_i^2+\gamma_i^{2}+\zeta_i^2)}{\tau}}\left(\frac{q'}{q}\right)^{\frac{17}{12}-\frac{1}{3}}\times\,\\
&&\qquad\qquad\left\{
\frac{1}
{\theta(w^{'-2};q')}\,
{\cal I'}^T(e^{-\frac{\pi i}{3}}q^{'-\sixth}\frac{w^{'\frac{1}{3}}}{r'},e^{\frac{\pi i}{3}}q^{'\sixth}\frac1{w^{'\frac{1}{3}}r'},{\bf x}',{\bf y}')+\right.\nonumber\\
&&\quad\qquad\qquad\left.\frac{1}
{\theta(w^{'2};q')}\,\,
{\cal I'}^T(e^{\frac{\pi i}{3}}q^{'\sixth}
\frac{w^{'\frac{1}{3}}}{r'},e^{-\frac{\pi i}{3}}q^{'-\sixth}
\frac1{w^{'\frac{1}{3}}r'},{\bf x}',{\bf y}')\right\}\,\nonumber\\
&&\qquad\qquad \equiv
 i\,e^{\frac{3\pi i \sum_{i=1}^N(\eta_i^2+\gamma_i^{2}+\zeta_i^2)}{\tau}}\left(\frac{q'}{q}\right)^{\frac{13}{12}}\,{\cal I}'_{E_6}({\bf z}',{\bf x}',{\bf y}';q')
\ee
Here $\zeta_i$ are the chemical potentials correspoding to fugacities $z_i$, {\it i.e.} $z_i=e^{2\pi i \zeta_i}$.
Thus we obtain
\be\label{E6mod1}
q^{\frac{13}{12}}\,{\cal I}_{E_6}({\bf z},{\bf x},{\bf y})=
e^{\frac{3\pi i \sum_{i=1}^3(\eta_i^2+\gamma_i^{2}+\zeta_i^2)}{\tau}}\;i\,{q'}^{\frac{13}{12}}\,{\cal I}'_{E_6}({\bf z}',{\bf x}',{\bf y}')
\ee
We again note that 
\be
\frac{13}{12}=\frac{c_{E_6}}{2}\,.
\ee The over-all phase dependence on the flavor chemical potentials 
is as if there are three triplets of hypermultiplets for
each one of the three $SU(3)$ flavor groups as expected. Note also that this modular transformation 
is consistent with~\eqref{boxed}: number of effective vector multiplets for $E_6$ SCFT is $n_V=5$, 
and the number of hypermultiplets is $n_H=16$.  Following from these properties modular transformations of all $A_2$ theories of class ${\cal S}$
 can be written as in~\eqref{boxed}. We should stress that the analysis of this section relies on Argyres-Seiberg duality relating the $E_6$ SCFT to 
a Lagrangian theory: a relation which implies~\eqref{e6dif}. In particular, the modularly transformed expression ${\cal I}'_{E_6}$ involves changing integration
cycles in the index of $N_f=6$ $SU(3)$ ${\cal N}=2$ SYM used in~\eqref{e6dif}: a procedure which is hard
 to phrase inherently for $E_6$ SCFT without making a  direct reference to Argyres-Seiberg duality.

\

\

\section{Discussion}\label{discsec}

Let us briefly summarize and discuss our results. In this note we have shown that the (modified) 
Schur index for all Lagrangian ${\cal N}=2$ superconformal theories with $A_{n}$ type gauge groups has
very simple and concrete  modular properties~\eqref{boxed}.
Moreover, we have also argued that same properties persist for non-Lagrangian SCFT with $E_6$ flavor symmetry making it plausible that the result can be generalized to other non-Lagrangian SCFTs.  

The modular properties of the modified Schur index for Lagrangian theories can be summarized as follows. 
Given a gauge theory we can write the index as a collection of contour integrals over the gauge groups.\footnote{
It is worth mentioning here that the index of any Lagrangian theory is given in terms of contour 
integrals of theta-functions and thus is a Fourier coefficient of those. 
Such Fourier coefficients are known to be related to Ramanujan's mock modular forms (see e.g.~\cite{zwegers}).
It would be very interesting to investigate further any possible 
relations between the index and mock modular forms. See {\it e.g.}~\cite{Dabholkar:2012nd} for a recent appearance of Mock modular forms
in counting problems in physics. 
We thank D.~Gaiotto for pointing this out to us.
} 
 The integrands are simple elliptic functions of the integration variables with the modular parameter 
defined by the only superconformal fugacity of this index $q$.\footnote{
One can argue that this statement is about gauge non-invariant quantities. However, the integrand
is just the index of a theory with a flavor  symmetry (which is gauged by performing the integral) with the 
addition of Haar measure and the index of the vector multiplet, and thus one can rephrase the statement
in terms of properties of such flavor symmetries. 
}
Performing the modular transformation   amounts to two operations.
First, we rewrite the integrands in terms of modularly transformed quantities the non-trivial functional
dependence of which on the (transformed) integration parameters stays the same. Second, 
the integration contours change from being around ${\cal A}$-cycle of the torus to being along 
the ${\cal B}$-cycle of the dual torus. Moreover, the conformal anomaly $c$ naturaly appears
in the modularly transformed expressions.  Although, this procedure is
very simple it is not optimal since it relies on a particular Lagrangian 
representation of the theory since it involves changing integration cycles corresponding to gauge symmetries. Ultimately one would like to be able to phrase the modular properties 
in a more invariant way, {\it e.g.} on par   with the duality invariant expressions of the index 
discussed in~\cite{Gadde:2011uv,Gadde:2011ik,GRR}. The explicit modular property~\eqref{E6mod1}
 of the $E_6$ SCFT discussed here has the structure of the general expression~\eqref{boxed}.
It is tempting thus to conjecture that the procedure summarized in this paragraph, and
culminating in~\eqref{boxed}, is true also for non-Lagrangian ${\cal N}=2$ superconformal theories; at least those
 which can be connected to Lagrangian ones by dualities.

In this note we have taken a Hamiltonian approach towards the computation of the index
as a trace over the states. However, some of the properties we discussed, {\it e.g.}
the fact that the integrands in the index computation are elliptic functions of the gauged chemical
potentials, involve analytical continuation of the parameters of the index taking us away from the physical trace interpretation. A more appropriate physical definition of the analytically continued expressions would be  through a partition function
on $S^3\times S^1$. Thus, it will
 be very beneficial to study the Schur index as such a partition 
function, using and extending the results
 of~\cite{Festuccia:2011ws,Dumitrescu:2012ha}. For example, 
one would like to understand in detail why gauge chemical potentials naturally live on the same torus as
the geometric torus defined by the two circles corresponding  to symmetris generated by $E$ and 
$j_2-j_1$. The modular properties of the Schur index for Lagrangian theories directly descend from the $SL(3,{\mathbb Z})$ properties of
 elliptic Gamma functions~\cite{SL3Za} discussed in the context of superconformal index in~\cite{Spiridonov:2012ww}. In appendix~\ref{sl3app} we show explicitly how 
that comes about.  However, going beyond Schur index by adding more superconformal fugacities
or considering ${\cal N}=1$ theories some of the simple structure is lost:
the integrands are not doubly periodic functions and the indices before and after modular transformations have functionally different form.\footnote{
On the other hand, the added complexity of the problem in these setups allows for new 
very interesting properties of the superconformal indices to emerge. For example, equality of 
indices of Seiberg dual pairs of theories can be linked to what is called total ellipticity
property of the integrals defining those indices~\cite{Spiridonov:2009za,Total1,Total2,Spiridonov:2012ww}.
Such properties trivialize in our setup.
} 
 In particular, it is natural to speculate that it is possible to understand 
the modular properties of the Schur index as a symmetry of the $S^3\times S^1$ partition function,
whereas the more general $SL(3,{\mathbb Z})$ structure relates partition functions on different manifolds.  We leave this question for future explorations.

\

\

\section*{Acknowledgments}

We would like to thank 
C.~Beem, T.~Dimofte, G.~Festuccia, A.~Gadde, D.~Gaiotto, Z.~Komargodski,
J.~Maldacena, L.~Rastelli, and N.~Seiberg
 for very useful discussions.
We are grateful to  KITP, ICTS Bangalore, Aspen center for physics, and the Simons Center 
for Geometry and Physics for hospitality during different stages of this project.
This research  was supported in part by NSF grant PHY-0969448. 
The  research at KITP was  supported in part by the NSF grant PHY-1125915 and the 
research at the Aspen center for physics was supported in part by NSF Grant 1066293 .

\

\appendix

\section{Modular properties of elliptic Gamma functions
and the Schur index}\label{sl3app}

In this appendix we discuss the relation of the modular transformation of the Schur index
studied in this note and the $SL(3;\,{\mathbb Z})$ transformations of the elliptic Gamma functions~\cite{SL3Za}.\footnote{See also~\cite{DSpir} for a disscussion of integrals of modularly transformed expressions.}
The {\it full}, {\it i.e.} depending on three superconformal fugacities~\cite{Gadde:2009kb,Gadde:2011uv}, modified index of a free hypermultiplet is given by
\be\label{fullind}
{\cal I}_H=\Gamma\left(-t^\half\,z^{\pm1};\,p,\,q\right)\,.
\ee The Schur index is obtained by setting $t=q$ and then the dependence on $p$ is lost. The elliptic Gamma function is given by
\be
\Gamma\left(z;\,p,\,q\right)=\prod_{m,n=0}^\infty\frac{1-p^{m+1}q^{n+1}z^{-1}}{1-p^{m}q^{n}z^{}}\,.
\ee
Defining
\be
&&z=e^{2\pi i\,u}\,,\qquad p=e^{2\pi i\,\tau}\,,\qquad q=e^{2\pi i\,\s}\,,\\
&&z'=e^{2\pi i\,u/\tau}\,,\qquad p'=e^{-2\pi i/\tau}\,,\qquad q'=e^{2\pi i\,\s/\tau}\,,\nonumber\\
&&z''=e^{2\pi i\,u/\s}\,,\qquad p''=e^{2\pi i\,\tau/\s}\,,\qquad q''=e^{-2\pi i/\s}\,,\nonumber
\ee the elliptic Gamma function has the following modular property~\cite{SL3Za}\footnote{
Here we assume that the parameters $\tau$ and $\s$ are such that both sides of the equality are well defined. If not, one can write a similar expression by exchanging all $p$ and $q$ variables.
}
\be\label{sl3z}
&&\frac{\Gamma\left(z'';\,p'',\,q''\right)}{\Gamma\left(z'/q';\,p',\,1/q'\right)}=e^{i\pi \,Q(u,\tau,\s)}\;\Gamma\left(z;\,p,\,q\right)\,,\\
&&Q(u,\tau,\s)=\frac{u^3}{3\tau\s}-\frac{\tau+\s-1}{2\tau\s}u^2+\frac{\tau^2+\s^2+3\tau\s-3\tau-3\s+1}{6\tau\s}u+\nonumber\\
&&\qquad\qquad\qquad+\frac{(\tau+\s-1)(\tau^{-1}+\s^{-1}-1)}{12}\,.\nonumber
\ee
 To reduce this relation to the modular property of the Schur index we want to set $t=q$.
Thus we obtain that
\be\label{sl3zH}
\frac{\Gamma\left(-{q''}^\half\, {z''}^{\pm1};\,p'',\,q''\right)}{\Gamma\left({(p'/q')}^\half {z'}^{\pm1};\,p',\,1/q'\right)}=e^{i\pi \,Q(-\half + \half\s\pm u ,\tau,\s)}\;\Gamma\left(-q^\half\,z^{\pm1};\,p,\,q\right)\,.
\ee Using the following identities
\be\label{props}
&&\Gamma\left(z;p,q\right)\;\Gamma\left(\frac{p\,q}{z};p,q\right)=1\,,\qquad
\Gamma\left(-q^\half z^{\pm1};p,q\right)=\frac{1}{\theta(-q^\half\,z;q)}\,,\\
&&Q(-\half + \half\s+ u ,\tau,\s)+Q(-\half + \half\s- u ,\tau,\s)=
\frac{1}{12}(\s+\s^{-1})-\frac{u^2}{\s}
\,,\nonumber
\ee  we can write~\eqref{sl3zH} as
\be
q^{\frac{1}{24}}\,\frac{1}{\theta(-q^\half\,z;q)}=e^{\frac{\pi i\,u^2}{\s}}{q''}^{\frac{1}{24}}\,\frac{1}{\theta(-{q''}^\half\,z'';q'')}\,,
\ee which coincides with~\eqref{freemod}.

\

Let us also consider setting $t=p$ in the $SL(3,{\mathbb Z})$ transformation~\eqref{sl3z} for the
free hypermultiplet. Here, we get
\be
\frac{\Gamma\left({(p''q'')}^\half\, {z''}^{\pm1};\,p'',\,q''\right)}{\Gamma\left(-{p'}^\half/{q'} {z'}^{\pm1};\,p',\,1/q'\right)}=e^{i\pi \,Q(-\half + \half\tau\pm u ,\tau,\s)}\;\Gamma\left(-p^\half\,z^{\pm1};\,p,\,q\right)\,,
\ee which can be again written using~\eqref{props} as
\be
p^{\frac{1}{24}}\,\frac{1}{\theta(-p^\half\,z;p)}=e^{\frac{\pi i\,u^2}{\tau}}{p'}^{\frac{1}{24}}\,\frac{1}{\theta(-{p'}^\half\,z';p')}\,.
\ee

\

Finally let us discuss another interesting limit of the index of a free hypermultiplet:
the Macdonald index~\cite{Gadde:2011uv}, $p=0$ or 
equivalently $\tau\to i\infty$. We parametrize 
\be
t=e^{2\pi i \rho}\,,
\ee and apply~\eqref{sl3z} to~\eqref{fullind} in this limit
\be
\lim_{\t\to i\infty}
\frac{e^{\frac{\pi i \tau (\s-\rho)}{6\s}}}
{\Gamma\left({(p't')}^\half/q' {z'}^{\pm1};\,p',\,1/q'\right)}=
\frac{e^{-\frac{\pi i\,u^2}{\s}}\,
\left(\frac{q}{q''}\right)^{\frac{1}{24}}\,
e^{-\frac{\pi i (\rho-\s)^2}{4\s}}
\;\Gamma\left(-t^\half\,z^{\pm1};\,0,\,q\right)}
{\Gamma\left((q''\,t'')^\half\, {z''}^{\pm1};\,0,\,q''\right)}\,.
\ee The limit on the left-hand-side has to be taken carefully since the variables $q'$, $p'$,
and $t'$ approach $1$ in the limit. Such a limit was worked out for instance in~\cite{Spiridonov:2010em},
\be
&&\lim_{\t\to i\infty} \Gamma\left(z';\,1/q',\,p'\right)\,e^{\frac{\pi i\,\tau}{12} (1+\s^{-1}+2\frac{u}{s}) }=
e^{\pi i P(u,\s)}\,
\frac{\left({z''}^{-1}q'';\,q''\right)}{\left({z}^{-1};\,q\right)}\,,\\
&&P(u,\s)=-\frac{1}{12}\left(\s+\s^{-1}+3(1+2u)+\frac{6}{\s}u(1+u)\right)\,.\nonumber
\ee
Using this relation we finally obtain the following transformation for the Macdonald index of the free
hypermultiplet,
\be
e^{\frac{2\pi i\,u^2}{\s}}\,e^{\frac{\pi i (\rho-\s)^2}{2\s}}\left(\frac{q}{q''}\right)^{-\frac{1}{12}}\,
\frac{\Gamma\left((q''\,t'')^\half\, {z''}^{\pm1};\,0,\,q''\right)
\left(-q/t^\half z^{\pm1};\,q\right)}{\left((q''/t'')^\half {z''}^{\pm1} ;\,q''\right)}
=
\Gamma\left(-t^\half\,z^{\pm1};\,0,\,q\right)\,.\nonumber\\
\ee This transformation reduces to the one for the Schur index by taking $t=q$ as expected.\footnote{Note that if $t=q$ then ${t''}^\half=-1$.} Moreover, by moving terms around this identity can be simply written as a modular transformation of theta-functions~\eqref{modP0},
\be\label{thetamac}
q^{\frac{1}{12}}\,\frac{1}{\theta(-t^\half\,z^{\pm1};q)}=e^{\frac{2\pi i\,\left(u^2+
\left(\frac{\rho-\s}{2}\right)^2\right)}{\s}}{q''}^{\frac{1}{12}}\,\frac{1}{\theta((q''t'')^\half\,{z''}^{\pm1};q'')}\,.
\ee Thus, for the Macdonald limit although the modular properties are just the usual $SL(2,{\mathbb Z})$
of the theta-function, the transformed index has quite different functional form from the index
of a free hypermultiplet.
In  the Hall-Littlewood limit~\cite{Gadde:2011uv}, $p=q=0$,  the modular transformations trivialize.

\

\section{Explicit expressions for the modified Schur index}\label{appInt}

Let us give several examples of  concrete expressions for the modified Schur index.
The index of a general conformal quiver of class ${\cal S}$ with
$SU(2)$ gauge groups can be given using the data of the underlying Riemann surface.
Employing the technology of orthogonal polynomials of~\cite{Gadde:2011uv,Gadde:2011ik}
an $SU(2)$ quiver with $\frak g$ loops and $s$ external lines has the following modified Schur index
\be\label{indexGS}
{\cal I}^T_{\frak g,s}(1;q)
&=&(q;q)^{2\frak g-2-2s}\,\sum_{\lambda=0}^\infty(-1)^{s\lambda}
\frac{(\lambda+1)^s\;q^{\half\lambda(2\frak g-2+s)}}{(1-q^{\lambda+1})^{2\frak g-2+s}}\,.
\ee Analogous formulae can be written also for quivers with higher rank gauge groups.
The sphere with three punctures is a free theory and indeed performing the sum in~\eqref{indexGS} we get
\be
{\cal I}^T_{0,3}(1;q)=\frac{1}{\theta(-q^\half;q)^4}\,.
\ee An example of an interacting theory
is torus with one puncture, {\it i.e.} ${\cal N}=4$ SYM with a decoupled
free hypermultiplet. Here~\eqref{indexGS} can be written as
\be
{\cal I}^T_{1,1}(1;q)=\frac{(q;q)^2}{\theta(-q^\half;q)^2}\,\oint\frac{dz}{4\pi i z}\frac{\theta(z^{\pm2};q)}
{\theta(-q^\half z^{\pm2};q)}=
\frac{1}{(q;q)^2}\sum_{\lambda=0}^\infty \frac{q^\lambda}{(1+q^{\half+\lambda})^2}=
\frac{\psi^{(1)}_q(\frac{1}{2}-\frac{1}{2\tau } )}{(2\pi \tau)^2 \, q^\half\,(q;q)^2}\,.\nonumber\\
\ee Here $\psi^{(1)}_q(z)$ is the first derivarive of q-digamma function $\psi_q(z)$.
In general the indices~\eqref{indexGS} are expressible in terms of (derivatives of) q-digamma 
function,
\be
\psi_q(z)=-\ln(1-q)+\ln q\sum_{\ell=0}^\infty\frac{q^{\ell+z}}{1-q^{\ell+z}}\,.
\ee  For example for genus ${\frak g}=2,3,4,\dots$ without punctures one gets
\be
&&{\cal I}^T_{\frak g,\,0}(1;q)=
\sum_{\lambda=0}^\infty\left[\frac{(q;q)^2\,q^{\lambda}}{(1-q^{\lambda+1})^2}\right]^{\frak g-1}\,,\\
&&{\cal I}^T_{2,\,0}(1;q)={-q^{-1}\,(q;q)^{2}}\;
\frac{\psi_q^{(1)}(1)}{(2\pi \tau)^{2}}\,,\qquad
{\cal I}^T_{3,\,0}(1;q)=\frac{q^{-2}\,(q;q)^{4}}
{3!}\;\left(
\frac{\psi_q^{(1)}(1)}{(2\pi \tau)^{2}}+\frac{\psi_q^{(3)}(1)}{(2\pi \tau)^{4}}
\right)\,,\nonumber\\
&&{\cal I}^T_{4,\,0}(1;q)=\frac{-q^{-3}\,(q;q)^{6}}
{5!}\;\left(4
\frac{\psi_q^{(1)}(1)}{(2\pi \tau)^{2}}+5\frac{\psi_q^{(3)}(1)}{(2\pi \tau)^{4}}+
\frac{\psi_q^{(5)}(1)}{(2\pi \tau)^{6}}
\right)\,,\qquad\cdots\nonumber
\ee

\

\section{An identity involving Schur polynomials and theta functions}\label{schurapp}
The Schur index of a sphere with two maximal and two minimal punctures is given by~\cite{Gadde:2011ik,Gadde:2011uv}
\be
{\cal I}(a,b,{\bf x},{\bf y};q)&=&
\frac{(q^3;q)^2}{(q;q)^{6}
(q^{\frac{3}{2}}a^{\pm3};q)
(q^{\frac{3}{2}}b^{\pm3};q)
\prod_{i\neq j}
(q\,x_i/x_j;q)
(q\,y_i/y_j;q)}\times\\
&&\quad\sum_{\cal R}\frac{1}{\left[{\text{dim}}_q{\cal R}\right]^2}\,\chi_{\cal R}({\bf x})\,\chi_{\cal R}({\bf y})\,
\chi_{\cal R}(q^\half\, a,\,q^{-\half}a,\,a^{-2})\,
\chi_{\cal R}(q^\half\, b,\,q^{-\half}b,\,b^{-2})\,.\nonumber
\ee The sum is over representations ${\cal R}$ of $SU(3)$  and $\chi_{\cal R}(\mathbf z)$ are Schur polynomials of $SU(3)$.
The representations of $SU(3)$ are labeled by two integers ${\cal R}=(R_1,R_2)$ with $R_1\geq R_2$.
The index of $E_6$ SCFT is given by~\cite{Gadde:2011ik,Gadde:2011uv}
\be
 {\cal I}_{E_6}({\bf z},{\bf x},{\bf y})=
\frac{(q^2;q)(q^3;q)}{(q;q)^{6}
\prod_{i\neq j}(q\,z_i/z_j;q)
(q\,x_i/x_j;q)
(q\,y_i/y_j;q)}\sum_{\cal R}\frac{1}{{\text{dim}}_q{\cal R}}\,\chi_{\cal R}({\bf x})
\,\chi_{\cal R}({\bf y})\,\chi_{\cal R}({\bf z})\,.\nonumber\\
\ee
The q-dimension is defined to be
\be
{\text{dim}}_q{\cal R}=
\chi_{\cal R}(q,q^{-1},1)=q^{-R_1}\frac{(1-q^{2+R_1})(1-q^{1+R_2})(1-q^{1+R_1-R_2})}{(1+q)(1-q)^3}\,.
\ee 
Thus~\eqref{e6dif} implies the following identity for arbitrary representation ${\cal R}$
of $SU(3)$,
\be
&&(1-q^2)\,{\text{dim}}_q{\cal R}
\,\frac{\chi_{\cal R}({\bf z})}{\prod_{i\neq j}(q\,z_i/z_j;q)}=\\
&&\qquad\qquad\left.\frac{
\chi_{\cal R}(q^\half\, a,\,q^{-\half}a,\,a^{-2})
\chi_{\cal R}(q^\half\, b,\,q^{-\half}b,\,b^{-2})}{\theta(w^{-2};q)(q^{\frac3{2}}a^{\pm3};q)
(q^{\frac32}b^{\pm3};q)}\right|_{a=q^{-\sixth}\frac{w^{\frac13}}{r},\;
b=q^{\sixth}\frac{w^{-\frac13}}{r}}
+\nonumber\\
&&\qquad\qquad\quad\left.\frac{\chi_{\cal R}(q^\half\, a,\,q^{-\half}a,\,a^{-2})\,
\chi_{\cal R}(q^\half\, b,\,q^{-\half}b,\,b^{-2})}{\theta(w^{2};q)(q^{\frac{3}{2}}a^{\pm3};q)
(q^{\frac{3}{2}}b^{\pm3};q)}\right|_{a=q^{\sixth}\frac{w^{\frac13}}{r},\,
b=q^{-\sixth}\frac{w^{-\frac13}}{r}}\,.\nonumber
\ee
Here ${\mathbf z}=\left(wr,w^{-1}r,r^{-2}\right)$.
This identity is the statement of Argyres-Seiberg duality for the Schur index.
The above formula can be viewed as splitting the $SU(3)$ maximal puncture into two minimal punctures.
In more general we would expect that a maximal puncture of $SU(N)$ can be ``split'' into $N-1$
minimal punctures using similar identities.

\

\

\

\bibliography{nonconf}

\bibliographystyle{JHEP}

\end{document}